\documentclass[twoside]{article}
\usepackage[latin1]{inputenc}
\usepackage{t1enc}
\usepackage{a4}
\usepackage{tabularx}
\usepackage{graphicx}
\usepackage{amssymb}
\usepackage{longtable}
\usepackage{natbib}
\bibpunct{(}{)}{;}{a}{}{,}

\makeatletter
\renewenvironment{thebibliography}[1]{%
 \subsection*{References}%
 \parskip=12pt
 \thebib@list
 \parskip=0pt
 \small
 \sloppy
}{%
 \endlist
}%
\def\thebib@list{%
  \list{\null}{%
  \leftmargin 3em\itemindent-\leftmargin
 }%
}%
\makeatother
\makeatletter
\renewcommand{\fps@figure}{thbp}
\renewcommand{\fps@table}{thbp}

\makeatother
\makeatletter
\def\LT@makecaption#1#2#3{%
  \LT@mcol\LT@cols c{\hbox to\z@{\hss\parbox[t]\LTcapwidth{%
    \sbox\@tempboxa{\normalsize #1{#2: }#3}%
    \ifdim\wd\@tempboxa>\hsize
      #1{#2: }#3%
    \else
      \hbox to\hsize{\hfil\box\@tempboxa\hfil}%
    \fi
    \endgraf\vskip\baselineskip}%
  \hss}}}
\makeatother

\newcommand{\kms}{{\,\rm km\,s}^{-1}} 
\newcommand{\arcsec}{\mbox{$.\!^{\prime\prime}$}}%

\newcommand\fm{\mbox{$.\!\!^{\mathrm m}$}}%

\newcommand\la{\mathrel{\hbox{\rlap{\hbox{\lower4pt\hbox{$\sim$}}}\hbox{$<$}}}}
\newcommand\ga{\mathrel{\hbox{\rlap{\hbox{\lower4pt\hbox{$\sim$}}}\hbox{$>$}}}}

\begin{document}

\setcounter{figure}{0}
\setcounter{table}{0}
\setcounter{section}{0}
\setcounter{equation}{0}

\begin{center}
{\Large\bf
The Ups and Downs of the Hubble Constant}\\[0.7cm] 

G.\,A. Tammann \\[0.17cm]
(in collaboration with B.~Reindl) \\[0.17cm]
Astronomisches Institut der Universit{\"a}t Basel \\
Venusstrasse 7, CH-4102 Binningen, Switzerland \\
G-A.Tammann@unibas.ch
\end{center}

\vspace{0.5cm}

\begin{abstract}
\noindent{\it
A brief history of the determination of the Hubble constant $H_{0}$ is
given. Early attempts following \citet{Lemaitre:27} gave much too high
values due to errors of the magnitude scale, Malmquist bias and
calibration problems. By 1962 most authors agreed that $75\la H_{0}
\la 130$. After 1975 a dichotomy arose with values near 100 and others
around 55. The former came from apparent-magnitude-limited samples and
were affected by Malmquist bias. New distance indicators were
introduced; they were sometimes claimed to yield high values of
$H_{0}$, but the most recent data lead to $H_{0}$ in the 60's, yet
with remaining difficulties as to the zero-point of the respective
distance indicators. SNe\,Ia with their large range and very small
luminosity dispersion (avoiding Malmquist bias) offer a unique
opportunity to determine the large-scale value of $H_{0}$. Their
maximum luminosity can be well calibrated from 10 SNe\,Ia in local
parent galaxies whose Cepheids have been observed with HST. An
unforeseen difficulty -- affecting {\em all} Cepheid distances -- is
that their P-L relation varies from galaxy to galaxy, presumably in
function of metallicity. A proposed solution is summarized here. The
conclusion is that $H_{0}=63.2\pm1.3$ (random) $\pm5.3$ (systematic)
{\em on all scales}. The expansion age becomes then (with $\Omega_{\rm
  m}=0.3, \Omega_{\Lambda}=0.7$) $15.1\;$Gyr.}
\end{abstract}

\section{Introduction}
\label{sec:01}
The present value of the Hubble parameter is generally called ``Hubble
Constant'' ($H_{0}$). The {\em present\/} value requires minimum
look-back-times; it is therefore to be determined at the smallest
feasable distances and is adequately defined by
\begin{equation}
H_{0}=\frac{v}{r}\;[\mbox{km\,s}^{-1}\;\mbox{Mpc}^{-1}],
\label{eq:01}
\end{equation}
where $v=cz$, $z=\Delta\lambda/\lambda_{0}$, and $r=$ distance in
Mpc. As long as $z\ll 1$, it is indicated to interprete $cz$ as a
recession velocity because the observer measures the {\em sum\/} of
the space expansion term $z_{\rm cosmic}={\cal R}_{0}/{\cal
R}_{\rm emission} -1$ (${\cal R}$ being the scale factor) and
$z_{\rm pec}$ caused by the density fluctuation-induced peculiar
motions. At small $z_{\rm cosmic}$ and in high-density regions
$z_{\rm pec}$ is not negligible. It is therefore mandatory to
measure $H_{0}$(cosmic) at distances where 
$z_{\rm cosmic}\gg z_{\rm pec}$ and outside of clusters. Any
determination of $H_{0}$ must therefore compromise between two
conditions: the smallest possible galaxy distances $r$ and a minimum
influence of $z_{\rm pec}$. The local Group is obviously useless
for the determination of $H_{0}$ because it is probably
gravitationally bound. The nearby Virgo cluster affects the local
expansion field out to $\sim\!2000\kms$ (see Section~\ref{sec:06}).
At $v\sim3000\kms$ the relative contribution of random velocities of
field galaxies decreases  to less than $10\%$, yet a volume of roughly
similar radius has a bulk motion of $630\kms$ with respect to the
CMB. To be on the safe side it is therefore desirable to trace
$H_{0}$ out to say $\sim\!20\,000\kms$.
The expansion rate at this distance is for all practical purposes
still undistinguishable from its present value.

     The first spectra of galaxies and the measurement of their radial
velocities by \citet{Slipher:14} and later by M.~Humason and others
was an epochal achievement. Today the observation of the 
redshifts needed for the calibration of $H_{0}$ is routine. The
emphasis here lies therefore entirely on the determination of galaxy
distances.

\section{The First Galaxy Distances}
\label{sec:02}
While the question as to the nature of the ``nebulae'' was still wide
open, \citet{Hertzsprung:14} applied the period-luminosity (P-L)
relation of Cepheids, which he had calibrated with Galactic Cepheids,
and found a distance modulus of SMC of $(m-M)_{\rm SMC}=20.3$
(115\,kpc), roughly a factor 1.8 too large. According to the custom of
the time he transformed the distance into a trigonometric parallax of
$0\arcsec0001$, losing a factor of 10 during the process. While
transforming the parallax into light years he lost another factor of
ten. Thus his published distance of 3000 light years buried his
sensational result. 

     In the following year \citet{Shapley:15} repeated Hertzsprung's
measurement. For various reasons he now obtained a Cepheid distance of
only $(m-M)_{\rm SMC}=16.1$ (17\,kpc), which he slightly increased in
\citeyear{Shapley:18} and which he could take as a confirmation of
his conviction that all ``nebulae'' were part of his very large
Galactic system. 

     \citet{Lundmark:20} was the first to recognize supernovae as a
class distinct from novae. This explained the brightness of the
``nova'' 1885 in M\,31 and led him to a modulus of 
$(m-M)_{\rm M31}=21.3$ (180\,kpc). Still much too low the value
could not be accommodated within even the wildest size estimates of
the Galaxy. But the result had no influence on the ``Great Debate''
\citep[cf.][]{Fernie:70}. 

     \citet{Oepik:21,Oepik:22} ingeniously used the rotation velocity
of M\,31 to determine the mass-to-light ratio of the galaxy and he
broke the distance degeneracy of this value by adopting a very
reasonable mass-to-light ratio of the Solar neighborhood. He obtained
a stunningly good value of $(m-M)_{\rm M31}=24.5$ (750\,kpc), which
he decreased in the following year by a factor of 1.7. {\"O}pik's
papers remained unnoticed. 

     The discovery of several novae in ``nebulae'', first by
\citet{Ritchey:17}, stimulated the search for variability and led
Hubble to the discovery of a Cepheid in M\,31 in 1923, -- the first
Cepheid beyond the Magellanic Clouds. At the meeting of the
Association for the Advancement of Science in December 1924 he
announced the discovery of several very faint Cepheids in M\,31. 
They proved that many of the nebulae are actually ``island
universes'', but the proof was not yet generally accepted, because van
Maanen's (\citeyear{vanMaanen:23}) claim of a detectable rotation of
the spirals. Hubble published his Cepheid distance of M\,31 only in
\citeyear{Hubble:29a}, after he had published the Cepheids in
NGC\,6822 (\citeyear{Hubble:25}) and  M\,33 (\citeyear{Hubble:26}).     

     Hubble used his Cepheid distances to calibrate the brightest
stars ($M_{\rm pg}=-6\fm3$) and the mean luminosity of ``bright''
galaxies ($M_{\rm pg}=-15\fm8$; either value being $4^{\rm m}-5^{\rm m}$ too
faint). In this way he extended his distance scale out to the Virgo
cluster. In \citeyear{Hubble:29b} he plotted 31 of his distances
against Slipher's radial velocities. Not without remaining doubts, he
concluded from the correlation of these two parameters that the
Universe was expanding and that the expansion rate was
$H_{0}=500$
-- a value which he never decisively revised. His paper is generally
considered to be the discovery of the expanding Universe, although
\citet{Lemaitre:27} and \citet{Robertson:28} had anticipated the
result and published expansion rates -- using Hubble's distances --
corresponding to $H_{0}=627$ and 461, respectively. 
\citet{deSitter:30} used 54 galaxy {\em diameters\/} and radial
velocities out to the Coma cluster -- again making extensive use of
Hubble's data -- to derive $H_{0}=461$. \citet{Oort:31}, questioning
Hubble's bell-shaped galaxian luminosity function and increasing the
luminosity of the really big galaxies, concluded that
$H_{0}\approx290$. The result was important (yet hardly noticed)
because it implied an expansion age of $\sim\!3.5\;$Gyr (For
$q_{0}=0$) and removed the open contradiction with geological ages of
the time.

     For the next 20 years little was done on $H_{0}$ until
\citet{Behr:51} challenged Hubble's value. He noticed the large
luminosity scatter of Local Group galaxies and he argued via the
Malmquist effect that Hubble's {\em mean\/} luminosity was too faint
by $\sim\!1\fm5$ if applied to more distant, 
magnitude-selected galaxies. (This is to my knowledge the first
mentioning of Malmquist statistics in extragalactic work). 
Citing \citet{Baade:44} he also corrected Hubble's magnitudes by
$0\fm35$ (at $18\fm3$). These were Behr's two main reasons for
deriving a value of $H_{0}=260$. He would have found an even smaller
value had he known of Stebbins, Whitford, \& Johnson's
(\citeyear{Stebbins:etal:50}) pioneering photoelectric photometry
which  proved Hubble's photometric scale error to be even larger. 

\begin{figure}[t]
\centering
\resizebox{0.525\textwidth}{!}{\includegraphics{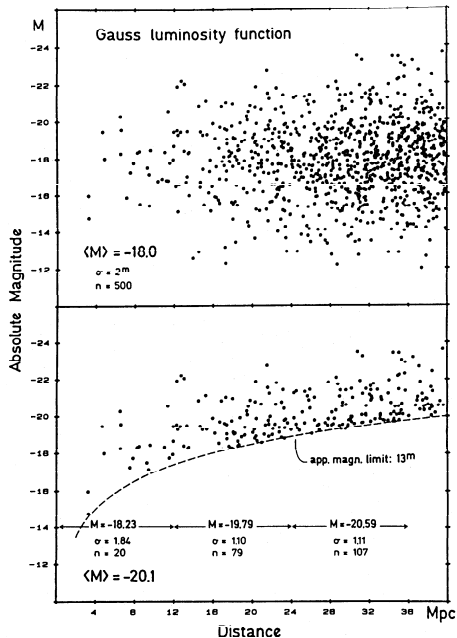}}
\caption{A Monte Carlo demonstration of the Malmquist Bias for 1000
  ``standard candles'' of fixed mean luminosity ($-18^{\rm m}$),
  non-zero luminosity dispersion ($\sigma=2^{\rm m}$) and
  $r<40\;$Mpc. Constant space density is assumed. 
  Upper panel: The unbiased distribution in absolute magnitude of a
  {\em distance-limited\/} sample. Lower panel: The same sample, but
  cut by an {\em apparent-magnitude\/} limit ($13^{\rm m}$). Note the
  increasing mean luminosity and decreasing magnitude dispersion in
  progressive distance intervals. (Only the magnitude dispersion of
  the entire sample in the lower panel happens to be close to the true
  dispersion in the upper panel). (By kindness of A.~Spaenhauer).}
\label{fig:01}
\end{figure}
The \citet{Malmquist:20,Malmquist:22} bias of
apparent-magnitude-limited samples as opposed to distance-limited
samples (which are very hard to come by) was fully acknowledged by
stellar astronomers since the 1920's, but it has beset -- if neglected
-- the extragalactic distance scale until quite recent times and led
consistently to too high values of $H_{0}$. The effect is illustrated
in Fig.~\ref{fig:01} and shows that in magnitude-limited
samples the mean absolute magnitudes of ``standard candles'' with
non-vanishing luminosity dispersion becomes brighter with increasing
distance. -- A smaller, but frequent overestimate of $H_{0}$ comes in
case of several individual determinations by averaging over $H_{i}$,
instead of over $\log H_{i}$.

     In later years many ways have been proposed how to correct
apparent-magnitude-limited samples in general and of field galaxies in
particular for Malmquist bias 
\citep[e.g.][for a tutorial see \protect{\citealt{Sandage:etal:95}}]%
{Spaenhauer:78,Tammann:etal:79,Teerikorpi:84,Teerikorpi:97,
  Bottinelli:etal:86,Sandage:96,Sandage:99b,Sandage:02,
  Theureau:etal:97,Goodwin:etal:97,Paturel:etal:98,Ekholm:etal:99,
  Butkevich:etal:05}.
Also cluster samples are affected by ``Teerikorpi Cluster Population
Incompleteness Bias'' \citep{Teerikorpi:87,Sandage:etal:95}. The hope
that the inverse TF relation was bias-free has not substantiated
\citep{Teerikorpi:etal:99}. In all cases the correction for Malmquist
  bias requires large and fair samples.

     \citet{Baade:48} had described the determination of improved
extragalactic distances as one of the major goals of the future
$200^{\prime\prime}$ telescope. Contrary to Behr he stirred anything short of
a sensation when he (\citeyear{Baade:52}) announced that work in M\,31
had shown, that either the zero-point of the Cepheids or of the
RR\,Lyr stars must be in error. 
Since Sandage's (published \citeyear{Sandage:53})
color-magnitude diagram of M\,3 had shown that the RR\,Lyr stars are
correct, the Cepheid luminosities had to be increased, as
\citet{Mineur:45} had already suggested. Baade concluded that
``previous estimates of extragalactic distances \dots were too small
by as much as a factor of 2'', which led him to $H_{0}\sim 250$. 
Accounting for the first four years of research with the $200^{\prime\prime}$
telescope, \citet{Sandage:54}, including also novae, summarized the
evidence for $H_0$ and concluded $125<H_{0}<276$.

\begin{figure}[t]
\centering
\resizebox{0.58\textwidth}{!}{\includegraphics{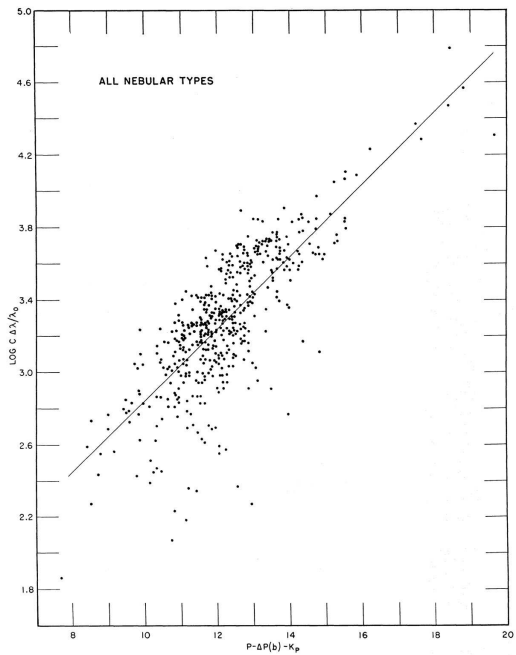}}
\caption{The Hubble diagram of the 474 field galaxies with redshifts
  known in 1956. The photographic magnitudes are corrected for
  Galactic absorption and the K-effect (due to redshift). The full
  line has slope 0.2 corresponding to linear expansion. A fit to the
  data gives a steeper slope, because the mean luminosity increases
  with distance due to Malmquist bias. \citep[From][]{Humason:etal:56}.}
\label{fig:02}
\end{figure}
   In their fundamental paper \citet*{Humason:etal:56} estimated $H_{0}=180$ on two grounds: 
(1) They showed that what Hubble had considered as brightest stars of
NGC\,4321, a member of the Virgo cluster, were actually HII
regions. The brightest stars set in only $\sim\!2^{\rm m}$ fainter. 
(2) The absolute magnitude of M\,31, resulting from its apparent
Cepheid modulus of $(m-M)=24.25$ \citep{Baade:Swope:54}, could be used
by the authors to calibrate the {\em upper-envelope line\/} of their
Hubble diagram of field galaxies on the assumption that the luminosity
of M\,31 must be matched by at least some galaxies. This elegantly
circumvented the problem of Malmquist bias (Fig.~\ref{fig:02}).

     The confusion between brightest stars and HII regions was
elaborated by \citet{Sandage:58}. The corresponding correction
together with the correction of Hubble's photometric scale led him to
conclude that the 1936 distance scale was too short by $4\fm6$ and
consequently that $H_{0}=75$. He noted that if the brightest stars had
$M_{\rm pg}=-9.5$ (which is now well demonstrated) $H_0$ would become
$55$. He also concluded from novae that Hubble's Local Group distances
were more nearly correct, i.e.\ too small by ``only'' $2\fm3$ on
average. Sandage's paper has become a classic for not only having
given the first modern values of $H_0$, but also because it contains
the first physical explanation of the instability strip of Cepheids.

\begin{figure}[t]
\centering
\resizebox{0.6\textwidth}{!}{\includegraphics{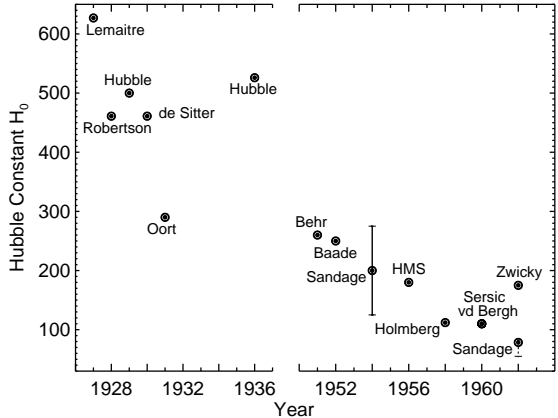}}
\caption{Determinations of $H_{0}$ from 1927 to 1962.}
\label{fig:03}
\end{figure}

     The situation in mid-1961 was summarized by \citet{Sandage:62} at
the influential 15th IAU Symposium in Santa Barbara. While he
cited values of $H_{0}\sim110$ by \citet{Sersic:60},
\citet{vandenBergh:60}, and \citet{Holmberg:58}, his own values --
based, in addition to Cepheids and brightest stars, on the {\em
  size\/} of HII regions -- were 75$-$82 and possibly as low as
55. F.~Zwicky pleaded in the discussion for $H_{0}=175$ from
supernovae. The decrease of $H_{0}$ from 1927 to 1962 is illustrated
in Fig.~\ref{fig:03}.

\section{Work on \boldmath{$H_{0}$} in 1962-1975}
\label{sec:03}
A new epoch began with the Cepheid distance of M\,31
of $(m-M)^{0}=24.20\pm0.14$ \citep{Baade:Swope:63}, derived by H.\,H.
Swope after W. Baade's death from his $200''$-plates 
and from H.\,C. Arp's photoelectric sequence. (For the history of the
time cf.\ also \citealt{Sandage:98,Sandage:99a}).

   By the same time the ``direct'' (i.e.\ non-spectroscopic) staff
members at the Mount Wilson and Palomar observatories (W. Baade,
E. Hubble, M. Humason, A. Sandage, and others) had accumulated many
$200''$-plates of a few galaxies outside the Local Group for work on
the Cepheids. Hubble and Baade had left their observations to Sandage,
who in addition had set up photoelectric sequences around these
galaxies, whose faintness and quality has remained  unsurpassed until
the advent of CCD detectors. Thus there was a unique wealth of
observations when I had the privilege to join the project as Sandage's
assistant in 1963.

   Although the first Cepheid distance of NGC\,2403 to come out of
the program confirmed Sandage's \citeyear{Sandage:62} value
\citep{Tammann:Sandage:68}, using the then latest version of the
Cepheid P-L relation \citep{Sandage:Tammann:68},
it was criticized as being (much) too large
\citep[e.g.][]{Madore:76,deVaucouleurs:78,Hanes:82}. The modern value
is actually only marginally smaller \citep{Saha:etal:05}.

   The second galaxy of the program, NGC\,5457 (M\,101), came as a
great surprise: its distance was found twice the value of Sandage's 
(\citeyear{Sandage:62}) estimate \citep{Sandage:Tammann:74a}, i.e.\
$(m-M)^{0}=29.3$. The distance of M\,101 and its companions was based on
brightest stars, HII region sizes, and van den Bergh's
(\citeyear{vandenBergh:60}) luminosity classes of spiral galaxies,
but also heavily on the {\em absence\/} of Cepheids down to the
detection limit. The faint Cepheids were eventually found with HST,
yielding  $(m-M)^{0}=29.34$ \citep{Kelson:etal:96} or $29.18$
\citep{Saha:etal:05}. In the mean time the distance had been denounced
as being too large 
\citep[e.g.][]{deVaucouleurs:78,Humphreys:Strom:83}. 

     The new distance of M\,101 made clear that the brightest spirals
of luminosity class (LC)\,I  are brighter than anticipated and
that the luminosity of their brightest stars and the size of their
largest HII regions had to be increased. This led immediately to a
distance of the Virgo cluster of $(m-M)=31.45$
\citep{Sandage:Tammann:74b}, -- a value probably only slightly too
small \citep[cf.][]{Tammann:etal:02}. The ensuing luminosity
calibration of LC\,I spirals could then be applied to a specially
selected, {\em distance-limited\/} sample of 36 such galaxies, bounded
by $8500\kms$. The conclusion was that $H_{0}=55\pm5$ ``everywhere''
\citep{Sandage:Tammann:75}. The largest contribution to the systematic
errors was attributed to the calibration through Cepheids.

\begin{figure}[t]
\centering
\resizebox{0.55\textwidth}{!}{\includegraphics{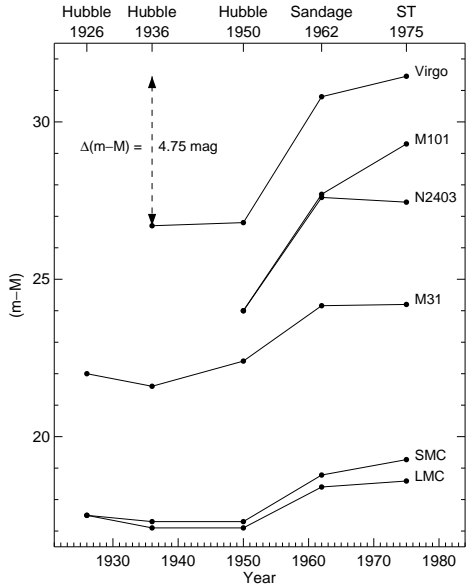}}
\caption{The development in time of some distances of local galaxies
  as stepping stones for the extragalactic distance scale. (Hubble's
  1950 distances in \citet{Holmberg:50}; ST stands for 
  Sandage \& Tammann).}
\label{fig:04}
\end{figure}

     In almost half a century from 1927 to 1975 the galaxy distances
have increased by roughly a factor of 10. The stretch factor is
non-linear, being $\sim\!2$ for the nearby LMC and SMC, but $\sim\!10$
for M\,101 and beyond (Fig.~\ref{fig:04}).

\section{\boldmath{$H_{0}$} after 1975}
\label{sec:04}
Work on $H_{0}$ exploded after 1975. The new activity was initiated by
G.~de Vaucouleurs.
Having started with $H_{0}=50$ from brightest globular
clusters \citep{deVaucouleurs:70}, he switched to $H_{0}\sim100\pm10$
\citep{deVaucouleurs:77,deVaucouleurs:Bollinger:79}. By assuming
rather short local distances and by turning a blind eye to all
selection effects, he managed to maintain this value -- eventually
with strong directional variations -- until his last paper on the
subject \citep{deVaucouleurs:Peters:85}.

     Old and new methods of distance determinations were
employed. They may be divided into 1) those using individual objects
in galaxies, and 2) those relying on global galaxian properties.

\subsection{Individual objects as distance indicators}
\label{sec:04:1}
\noindent
 {\em a) RR Lyr stars.} Extensive work on their luminosity
   calibration in function of metallicity seems now to converge, but
   their remain some exceptions. Their range is so far confined to the
   Local Group. For a review see \citet{Sandage:Tammann:06}.

\noindent
 {\em b) Cepheids.} See Section~\ref{sec:05:2:2} and
   \citet{Sandage:Tammann:06}. 

\noindent
 {\em c) Brightest stars.} The luminosity of brightest stars,
   Hubble's classical vehicle, lost much of its grip when it was shown
   that it depends on the size (luminosity) of the parent galaxy
   \citep{Sandage:Tammann:74aa}.

\noindent
 {\em d) Size of HII regions.} The size of the largest HII regions
   in late-type galaxies was introduced as a distance indicator by
   \citet{Sersic:60} and \citet{Sandage:62}. Imaging of many galaxies
   with an H$\alpha$ filter by Sandage extended the distance scale
   considerably \citep{Sandage:Tammann:74a}, but also here it was
   found that the size depends on the size of the parent galaxy. The
   method is not competitive anymore.

\noindent
 {\em e) Globular clusters (GCs).} The luminosity of the peak of the
   bell-shaped luminosity function (LF) of GCs has been proposed as
   a standard candle \citep{vandenBergh:etal:85}. The method seems
   attractive because its calibration depends on the well defined LF
   of Galactic GCs whose Population~II distances are independent of
   Cepheids. It was employed by several authors 
   \citep[for reviews see][]{Harris:91,Whitmore:97,Tammann:Sandage:99}.  
   But the basic {\em assumption\/} that the LF was universal is
   shattered by the fact that some GC color functions and LFs show
   double peaks, and by doubts that the formation of GCs is a unique
   process.   

\noindent
 {\em f) Novae.} After the confusion of novae and supernovae had
   been lifted by \citet{Lundmark:20}, novae played a role as distance
   indicators in their own right. Instead of the luminosity at
   maximum, which has a very wide dispersion, the magnitude 15 days
   after maximum or the luminosity-decline rate relation were
   used. The independent calibration can come, at least in principle,
   from expansion parallaxes of Galactic novae \citep{Cohen:85}. The
   data acquisition of novae is demanding on telescope time and little
   has been done in recent years.

\noindent
 {\em g) Planetary nebulae (PNe).} Following a proposal by
   \citet{Ford:Jenner:78} also brightest planetary nebulae have been
   widely used as distance indicators. But the method seems to depend
   on population size \citep{Bottinelli:etal:91,Tammann:93}, chemical
   composition, and age \citep{Mendez:etal:93}; moreover the PNe in
   NGC\,4697 have different LFs depending on their dynamics
   \citep{Sambhus:etal:05}.

\noindent
 {\em h) The tip of the red-giant branch (TRGB).} It was shown by
   \citet{DaCosta:Armandroff:90} that the TRGB in globular clusters has
   a fixed absolute $I$-magnitude, irrespective of metallicity. The
   TRGB has hence been used as a distance indicator by several authors
   \citep{Lee:etal:93,Salaris:Cassisi:97,Madore:etal:97,Sakai:99,
   Karachentsev:etal:03,Sakai:etal:04}.  
   The method is of great interest since its calibration rests on
   Population~II objects (GCs and RR\,Lyr stars) and provides an
   independent test of the Cepheid distance scale. I will return to
   the point in Section~\ref{sec:05:2:2}.

\noindent
 {\em i) Supernovae of type Ia (SNe\,Ia).} See Section~\ref{sec:05:2:1}.

\subsection{Global properties of galaxies as distance indicators}
\label{sec:04:2}
\noindent
 {\em a) Luminosity classes (LC) of spiral galaxies.} The
   luminosity of a spiral galaxy correlates with the ``beauty'' of its
   spiral structure. Correspondingly they were divided into 
   class I (the brightest) to V (the faintest) by 
   \citet{vandenBergh:60b,vandenBergh:60c,vandenBergh:60d} with
   additional galaxies classified by \citet{Sandage:Tammann:81} and
   others. The purely morphological LC classification is independent
   of distance; it yields therefore relative distances which were
   valuable for many years when velocity distances where suspected to
   be severely distorted by peculiar and streaming motions. Locally
   calibrated LC\,I spirals out to $6000\kms$ from a distance-limited
   sample were used to derive $H_{0}=56.9\pm3.4$
   \citep{Sandage:Tammann:75}. Bias-corrected LC distances led
   \citet{Sandage:99b} to $H_{0}=55\pm3$.

\begin{figure}[t]
\centering
\resizebox{0.9\textwidth}{!}{\includegraphics{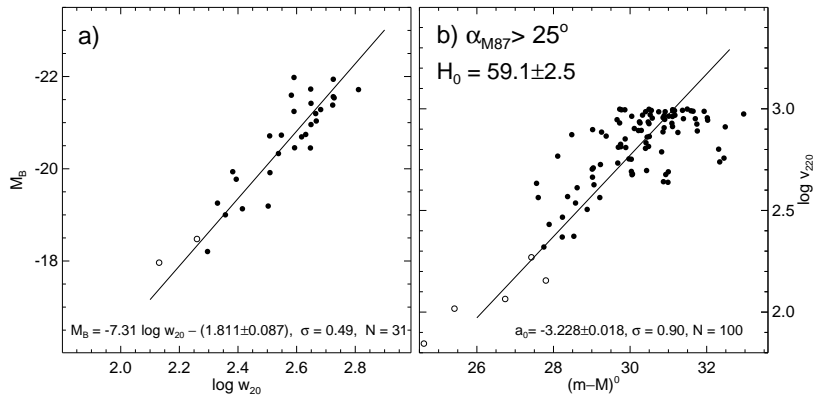}}
\caption{a) Calibration of the TF relation by means of 31 galaxies
  ($i>45^{\circ}$) with known Cepheid distances. The two open circles
  are companions to M\,101 which are assumed to be at the distance of
  M\,101. The slope is taken from the Virgo cluster. $w_{20}$ is the
  inclination-corrected line width at the 20\% intensity level
  expressed in $\kms$. b) The Hubble diagram of 100 field galaxies
  within $v_{220}<1000\kms$ and known TF distances. The turbulent
  region with a radius of $25^{\circ}$ about the Virgo cluster is
  omitted. Note the large scatter.}
\label{fig:05}
\end{figure}

\noindent
 {\em b) 21cm-line widths.} 21cm (or alternatively optical; see
   \citealt{Mathewson:etal:92,Mathewson:Ford:96})
   spectral line widths are a measure of a galaxy's rotation velocity,
   if corrected for inclination $i$, and hence correlate with its mass
   and luminosity \citep{Gouguenheim:69}. The relation was applied for
   distance determinations by \citet[][TF]{Tully:Fisher:77} and many
   subsequent authors, some of which are listed in
   Table~\ref{tab:02}. Several of the solutions for $H_{0}$ were
   dominated by Malmquist bias.
   The present (2005) calibration of the TF relation rests on 31
   galaxies with $i>45^{\circ}$ and with known Cepheid distances from
   \citet{Saha:etal:05}; the slope of the relation is taken from a
   complete sample of 49 inclined spirals in the Virgo Cluster
   (Fig.~\ref{fig:05}a). The scatter of $\sigma_{\rm m}=0\fm49$ is much
   larger than can be accounted for by errors of the Cepheid
   distances; it reflects mainly the large intrinsic scatter of the TF
   relation. 

     The calibration can be applied to an almost complete, {\em
   distance\/}-limited sample, as compiled by \citet{Federspiel:99},
   of 100 inclined spirals with $v_{220}<1000\kms$ (for the corrected
   velocities $v_{220}$ see Section~\ref{sec:06}). 
   The result in Fig.~\ref{fig:05}b gives $H_{0}=59.1\pm2.5$, but it
   is disappointing as to the very large scatter, which is much larger
   than from the calibration in Fig.~\ref{fig:05}a, even if the
   turbulent region of radius $25^{\circ}$ about the Virgo cluster is
   omitted. The reason is unclear; it cannot be due to peculiar
   motions which are much too small (see Section~\ref{sec:06}). It may
   be that remaining observational errors of the galaxian parameters
   contribute to the scatter. In any case the example shows that the
   TF method is difficult to handle. Open questions remain as to the
   large corrections for internal absorption, to truncated galaxies
   and hence to environment, and to the dependence on color and Hubble
   type. If only the apparently brightest galaxies were considered,
   arbitrarily large values of $H_{0}$ would be the consequence
   \citep[see][Fig.~4]{Tammann:etal:02}. The TF is therefore
   vulnerable against Malmquist bias, even if the intrinsic scatter
   was ``only'' $0\fm49$ as in Fig.~\ref{fig:05}a. 

     An application of the TF calibration in Fig.~\ref{fig:05}a to a
   {\em complete\/} sample of 49 untruncated spirals with
   $i>45^{\circ}$ in the Virgo cluster \citep{Federspiel:etal:98}
   yields a cluster modulus of $(m-M)^{0}=31.62\pm0.16$. Again,
   arbitrarily small values of the distance will emerge if the cluster
   sample is cut by an apparent-magnitude limit
   \citep[][Fig.~6]{Kraan-Korteweg:etal:88}.

\begin{figure}[t]
\centering
\resizebox{0.6\textwidth}{!}{\includegraphics{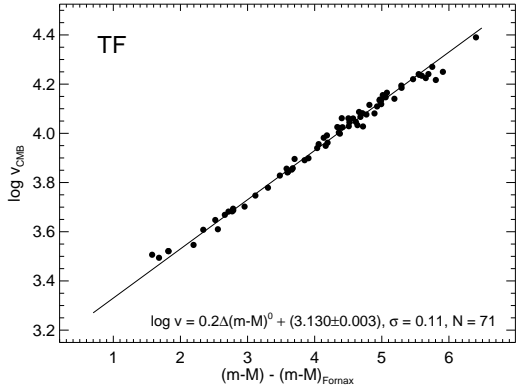}}
\caption{The Hubble diagram of 71 clusters whose distances relative to
  the Fornax cluster are known from the Tully-Fisher relation. Each
  point is the mean of about 10 galaxies. (Data from
  \citealt{Giovanelli:etal:97} and \citealt{Dale:etal:99}).} 
\label{fig:06}
\end{figure}

     \citet{Giovanelli:etal:97} and \citet{Dale:etal:99} have
   determined TF data for roughly 10 galaxies in each of 51 clusters
   with $3000<v_{\rm CMB}<25\,000\kms$. They define a Hubble line
   (Fig.~\ref{fig:06}) with very small scatter of $0\fm11$, which is even
   competitive with SNe\,Ia. Unfortunately the TF calibration of
   Fig.~\ref{fig:05}a cannot be directly applied to the cluster
   sample, because it cannot be assumed that the individual cluster
   galaxies were selected in the same way as the local galaxies for
   which Cepheid distances are available; this is a critical condition
   considering the large intrinsic dispersion if $\sigma\ga0\fm4$ of
   the TF method \citep[for an opposite view
   see][]{Giovanelli:97,Sakai:etal:00}. Instead it is possible to
   relate all cluster moduli to the modulus of the Fornax cluster. The
   ensuing equation of the Hubble line is shown at the bottom of
   Fig.~\ref{fig:06}. By simple transformation it follows
   \begin{equation}
     H_{0}= -0.2(m-M)_{\rm Fornax} + (8.130\pm0.003).
   \label{eq:02}
   \end{equation}
   Inserting the Fornax cluster modulus of $(m-M)^{0}=31.54\pm0.13$
   from Cepheids and SNe\,Ia (see below) leads to $H_{0}=65.6\pm4.1$.
     
   While the distance indicators under \ref{sec:04:2}\,a),\,b) involve spiral
   galaxies, the following three methods use E/S0 galaxies. The
   disadvantage is that there are no Cepheid distances -- nor RR~Lyr
   star or TRGB distances -- available for normal early-type galaxies
   to set the zero-point of the distance scale. In some cases one may
   infer an association between an E/S0 galaxy and a spiral with known
   Cepheid distance, or one may assume that the specific method
   applies also to the bulges of spiral galaxies. But these cases
   remain few, while actually many calibrators would be needed in view
   of the large intrinsic scatter of $\ga0\fm3$, i.e.\ much larger than
   that of SNe\,Ia.

\noindent
 {\em c) Brightest cluster galaxies (BCG).} The important potential
   of BCGs as standard candles to trace the expansion of the Universe
   was exploited first by Humason (\citeyear{Humason:36};
   \citealt{Humason:etal:56}). The work was propelled by
   \citet{Sandage:67,Sandage:68,Sandage:72,Sandage:73}. His papers
   were the decisive proof for cosmic expansion at a time when many
   astronomers speculated in view of the large quasar redshifts about
   a mysterious origin of redshifts. The last paper
   \citep{Sandage:Hardy:73} on the subject containing galaxies of
   moderate redshift lists 72 BCGs with $3500<v_{\rm
     CMB}<30\,000\kms$. They define a Hubble line of 
   \begin{equation}
     \log v = 0.2m_{\rm 1st} + (1.364\pm0.007)
   \label{eq:03}
   \end{equation}
   with a scatter of $\sigma_{\rm m}=0\fm29$. This implies
   \begin{equation}
     \log H_{0}= 0.2M_{\rm 1st} + (6.364\pm0.007).
   \label{eq:04}
   \end{equation}
   The mean absolute magnitude (in their corrected photometric system)
   of the two BCGs in the Virgo and Fornax clusters is $M_{\rm
   1st}=-23\fm15$, using the cluster distances from Cepheids and
   SNe\,Ia (see below). Hence $H_{0}=54.2\pm5.4$.

\noindent
 {\em d) The D$_{n}$-$\sigma$ or fundamental plane method (FP).} The
   correlation of the velocity dispersion $\sigma$ of E/S0 galaxies
   with their luminosity was pointed out by \citet{Minkowski:62} and
   \citet{Faber:Jackson:76}. Later the luminosity was replaced by a
   suitably normalized diameter D$_{n}$ \citep{Dressler:etal:87} or by
   surface brightness \citep{Djorgovski:Davis:87}. The method was
   extended to the bulges of spiral galaxies by \citet{Dressler:87}
   who derived $H_{0}=67\pm10$. \citet{Federspiel:99} used the great
   wealth of D$_{n}$-$\sigma$ data by \citet{Faber:etal:89} in two
   ways. First he derived the modulus difference between the Virgo and
   Coma cluster to be $3.75\pm0.20$ from 23 Virgo and 33 Coma
   members. With a Virgo modulus of 
   $(m-M)^{0}_{\rm Virgo}=31.47\pm0.16$ from Section~\ref{sec:06}
   one obtains therefore $(m-M)^{0}_{\rm
   Coma}=35.22\pm0.26$. Secondly he used an
   apparent-magnitude-limited subset of 264 early-type, 
   high-quality field and cluster galaxies brighter than $13\fm5$ to
   derive a value of $H_{0}$ after correcting for Malmquist bias
   following the method outlined in \citet{Federspiel:etal:94}. Beyond
   $v_{\rm CMB}=4000\kms$ his bias corrections become unreliable
   because the sample is far from being complete to the
   apparent-magnitude limit. That Malmquist bias must indeed be a
   major problem for the D$_{n}$-$\sigma$ and FP methods stems from
   their intrinsic scatter of $\sigma_{\rm m}=0\fm36$ as seen in the
   Coma \citep{Federspiel:99} and other cluster
   \citep{Joergensen:etal:96}.  
   For that reason claims of detected streamings toward the ``Great
   Attractor'' just outside $4000\kms$ \citep{Lynden-Bell:etal:88} are
   not beyond doubt. -- Within $v_{\rm CMB}=4000\kms$ Federspiel's
   (\citeyear{Federspiel:99}) analysis yields $H_{0}=57.0\pm4.4$ if
   the Virgo modulus from Section~\ref{sec:06} is adopted for the
   calibration. 

\begin{figure}[t]
\centering
\resizebox{0.6\textwidth}{!}{\includegraphics{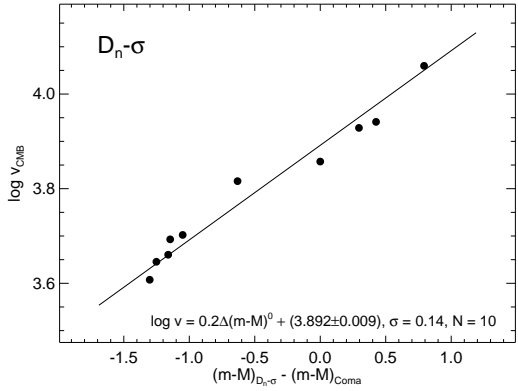}}
\caption{The D$_{n}$-$\sigma$ distances of 9 clusters relative to the
  Coma cluster. Each point is the mean of about 20 galaxies. (Data
  from \citealt{Joergensen:etal:96}).}
\label{fig:07}
\end{figure}

     \citet{Joergensen:etal:96} have gained D$_{n}$-$\sigma$ and FP
   observations of 232 E/S0 galaxies in 10 clusters and determined
   their mean distances relative to the Coma cluster. Their Hubble
   diagram is shown in Fig.~\ref{fig:07} for the D$_{n}$-$\sigma$
   distances, which have a slightly smaller scatter of
   $\sigma_{\rm m}=0\fm14$ than their FP
   distances. The scatter of $0\fm14$ about the Hubble line of the 9
   clusters beyond $v_{\rm CMB}=3700\kms$ is significantly larger than
   of SNe\,Ia ($\sigma_{\rm m}=0\fm10$, see \ref{sec:05:2:1}) and
   cannot mainly be explained by peculiar motions. The Hubble line in
   Fig.~\ref{fig:07} implies
   \begin{equation}
     \log H_{0}= -0.2(m-M)^{0}_{\rm Coma} + (8.892\pm0.009),
   \label{eq:05}
   \end{equation}
   which yields with $(m-M)^{0}_{\rm Coma}$ from above
   $H_{0}=70.4\pm9.0$. 

     An interesting by-product of the equation at the bottom of
   Fig.~\ref{fig:07} is the mean recession velocity of the
   Coma cluster freed from all peculiar velocities and streamings, if
   one assumes that the peculiar motions of the 9 clusters beyond
   $3700\kms$ average out. The zero-point of the relative distance
   scale should be reliable to within $0\fm05$ because 
   44 D$_{n}$-$\sigma$ distances are available for Coma. 
   From this follows an unperturbed velocity of 
   $v_{\rm Coma}=7800\pm200\kms$. 

\begin{figure}[t]
\centering
\resizebox{0.6\textwidth}{!}{\includegraphics{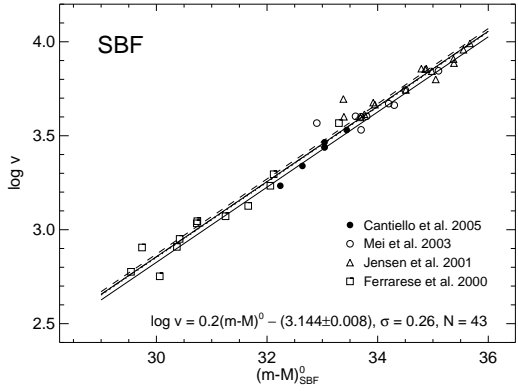}}
\caption{The Hubble diagram with SBF distances from different
  authors. Objects in the turbulent region within $25^{\circ}$ from
  the Virgo cluster center are not shown. The mean Hubble line
  suggests $H_{0}=71.8$, but the zero-point calibration remains
  unreliable.} 
\label{fig:08}
\end{figure}

\noindent
 {\em e) Surface brightness fluctuations (SBF).} 
   This method has been introduced by \citet{Tonry:Schneider:88} 
   and extensively used for E/S0 galaxies \citep{Tonry:etal:01}. 
   The size of the fluctuations shows little dependence on metallicity
   if measured in the infrared; the dependence on stellar population
   is compensated by allowing for the color $(V\!-\!I)$. 
   SBF distances of four recent investigations, based on observations
   with HST, are plotted in a Hubble diagram in Fig.~\ref{fig:08}. The
   distance zero-point depends entirely on Cepheid distances, either
   of up to six spirals whose bulges are treated like an E/S0 galaxy
   or/and of 1-5 aggregates containing E/S0's as well as spirals with
   Cepheid distances. The relative small scatter of
   $\sigma_{\rm m}=0\fm26$ beyond $3000\kms$, -- yet significantly
   larger than for SNe\,Ia in dust-free parent galaxies 
   ($\sigma_{\rm m}=0\fm10$, see \ref{sec:05:2:1}), 
   -- shows that the method works in principle. 
   But the result on $H_{0}$ is paradoxical being
   15\% larger even locally than from the direct evidence from
   Cepheids (see Sec.~\ref{sec:06}, Fig.~\ref{fig:12}b). If
   Cepheids are trusted at all, a Cepheid-calibrated distance
   indicator (SBF) must on average reproduce the distance scale of the
   Cepheids. There remains therefore a problem with the zero-point
   calibration; either because the bulges of spirals have different
   stellar populations as E/S0's, a possibility pointed out by
   \citet{Ferrarese:etal:00}, or because of unaccounted dust in spiral
   bulges. -- At larger distances also Malmquist bias may play a
   r{\^o}le in view of $\sigma_{\rm m}=0\fm26$. Many of the distant
   galaxies are brightest cluster galaxies, and it may not be
   warranted to extrapolate the SBF-magnitude relation from local
   calibrators -- some of them being only spiral bulges! -- to
   galaxies with $M\la -23\fm0$. The conclusion is that SBFs yield
   {\em relative\/} distances within $\sim\!13\%$, but that they are
   not (yet) to be used for the determination of $H_{0}$. The Coma
   cluster is not useful for the calibration at large distance
   because only three member galaxies have SBF measurement and the
   cluster distance itself has a large error.

     An overview of the present determinations of $H_{0}$ outside
   $v_{\rm CMB}>3000\kms$ is given in Table~\ref{tab:01}; the
   value from SNe\,Ia from \ref{sec:05:2:3} is anticipated for
   comparison. 
\begin{table}
\begin{center}
\caption{Present determinations of $H_{0}$ reaching out to $v_{\rm
   CMB}>3000\kms$.} 
\label{tab:01}
\footnotesize
\begin{tabular}{p{0.25cm}lcrlc}
\hline
\hline
\noalign{\smallskip}
   &
   Method &
   $\sigma_{\rm m}$ &
   \multicolumn{1}{c}{range} &
   \multicolumn{1}{c}{calibration$^{1)}$} &
   \multicolumn{1}{c}{$H_{0}\;^{2)}$} \\
    &
    intrinsic &
    \multicolumn{1}{c}{$\kms$} &
    &
   \\
\noalign{\smallskip}
\hline
\noalign{\smallskip}
& TF               & 0.45 & 25\,000 & Fornax                & $65.6\pm4.1$ \\
& BCG              & 0.30 & 30\,000 & Virgo + Fornax        & $54.2\pm5.4$ \\
& D$_{n}$-$\sigma$ & 0.36 &  4\,000 & Virgo                 & $57.0\pm4.4$ \\
& D$_{n}$-$\sigma$ & 0.36 & 10\,000 & Coma                  & $70.4\pm9.0$ \\
& SBF              & 0.26 & 10\,000 & Cepheid dist.         & ($71.6$) \\ 
& SNe\,Ia          & 0.10 & 30\,000 & 10 Cepheid dist.      & $62.3\pm1.3$ \\
\noalign{\smallskip}
\hline
\noalign{\smallskip}
\multicolumn{6}{l}{\small $^{1)}$ For easier comparison all underlying Cepheid distances
  are taken from }\\
\multicolumn{6}{l}{\small \citet{Saha:etal:05}} \\
\multicolumn{6}{l}{\small $^{2)}$ The systematic error of the Cepheid
  distance scale is not included}\\
\end{tabular}
\end{center}
\end{table}

     The mean value of $H_{0}=60.7\pm2.5$ from the first five lines in
Table~\ref{tab:01} is in statistical agreement with the value
from SNe\,Ia in \ref{sec:05:2:3}. The latter have the decisive
advantage of having very small scatter ($0\fm10$) and being hence
insensitive to Malmquist bias; it rests in addition on a solid
zero-point from 10 direct Cepheid distances.

\subsection{Various determinations of \boldmath{$H_{0}$} after 1975}
\label{sec:04:3}
\begin{figure}[t]
\centering
\resizebox{0.6\textwidth}{!}{\includegraphics{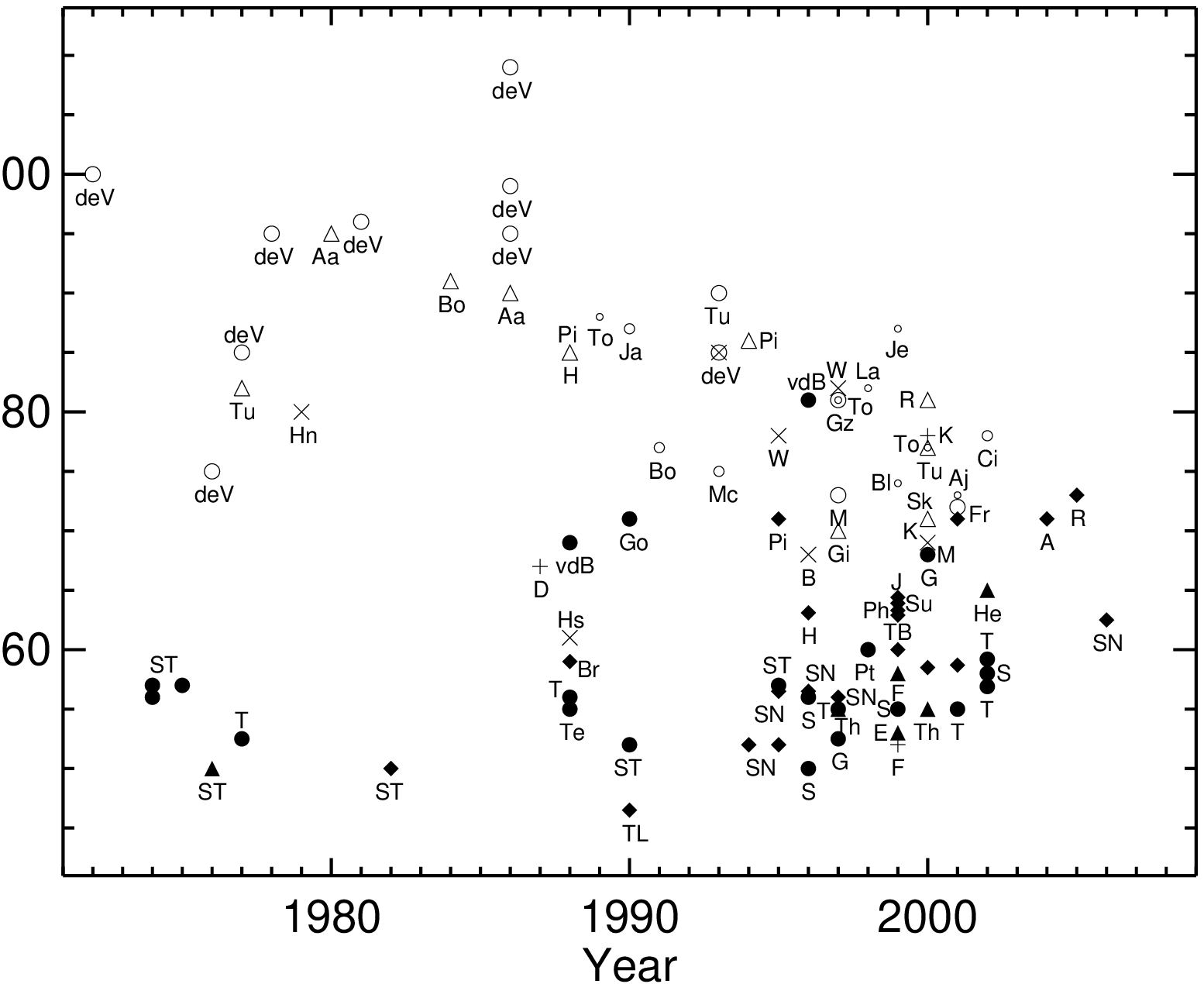}}
\caption{Various values of $H_{0}$ since 1975. Different symbols
  indicate different methods of distance determinations. Open symbols
  indicate when $H_{0}$ is based on apparent-magnitude-limited
  samples; closed symbols stand for bias-free or bias-corrected samples.}
\label{fig:09}
\end{figure}

The above distance indicators have been used in various
combinations to derive values of $H_{0}$. Many authors have
contributed; a representive subset has been compiled in
Table~\ref{tab:02}. The resulting values of $H_{0}$ since 1975 are
plotted in Fig.~\ref{fig:09} against the year of publication.

\footnotesize
\begin{longtable}{lllp{8.3cm}}
\caption{Values of $H_{0}$ from 1974 $-$ 2005\label{tab:02}.}\\[-8pt]
\hline
\hline
\noalign{\smallskip}
   Year &
   $H_{0}$ &
   Code &
   Reference \\
\noalign{\smallskip}
\hline
\endfirsthead
\caption{(Continued)}\\[-8pt]
\hline
\hline
\noalign{\smallskip}
   Year &
   $H_{0}$ &
   Code &
   Reference \\
\noalign{\smallskip}
\hline
\endhead
\noalign{\smallskip}
\hline
\endfoot
\noalign{\smallskip}
\hline
\endlastfoot
\noalign{\smallskip}
   \multicolumn{4}{c}{(a) various methods,  corr. for bias [$\bullet$]}\\
\noalign{\smallskip}
\hline
1974 & 56   & ST  & Sandage, A., \& Tammann, G.\,A. 1974, ApJ 194, 223 \\
1974 & 57   & ST  & Sandage, A., \& Tammann, G.\,A. 1974, ApJ 194, 559 \\
1975 & 57   & ST  & Sandage, A., \& Tammann, G.\,A. 1975, ApJ 196, 313; 197, 265 \\
1977 & 52.5 & T   & Tammann, G.\,A. 1977, in Redshifts and the Expansion of the Universe, 43 \\
1982 & 50   & ST  & Sandage, A., \& Tammann, G.\,A. 1982, ApJ 256, 339 \\
1988 & 69   & vdB & van den Bergh, S. 1988, in The Extragalactic Distance Scale, ASP Conf. Ser. 4, 375 \\
1988 & 56   & T   & Tammann, G.\,A. 1988, in The Extragalactic Distance Scale, ASP Conf. Ser. 4, 282 \\
1988 & 55   & Te  & Terndrup, D.\,M. 1988, in The Extragalactic Distance Scale, ASP Conf. Ser. 4, 211 \\
1990 & 71   & Go  & Gouguenheim, L., et al. 1990, in Proc. XXIVth Moriond Meeting, 3 \\
1990 & 52   & ST  & Sandage, A., \& Tammann, G.\,A. 1990, ApJ 365, 1 \\
1995 & 57   & ST  & Sandage, A., \& Tammann, G.\,A. 1995, ApJ 446, 1 \\
1996 & 56   & S   & Sandage, A. 1996, AJ 111, 1 \\
1996 & 50   & S   & Sandage, A. 1996, AJ 111, 18 \\
1996 & 81   & vdB & van den Bergh, S. 1996, PASP 108, 1091 \\
1997 & 52.5 & G   & Goodwin, S.\,P., Gribbin, J., \& Hendry, M.\,A. 1997, AJ 114, 2212 \\
1997 & 55   & T   & Tammann, G.\,A., \& Federspiel, M. 1997, in The Extragalactic Distance Scale, ed. M. Livio (Cambridge Univ. Press), 137 \\
1998 & 60   & Pt  & Paturel, G., et al. 1998, A\&A 339, 671 \\
1999 & 55   & S   & Sandage, A., 1999, ApJ 527, 479 \\
2000 & 68   & M   & Mould, J.\,R., et~al. 2000, ApJ 529, 786 \\
2001 & 55   & T   & Tammann, G.\,A., Reindl, B., \& Thim, F. 2001, in Cosmology and Particle Physics, AIP Conf. Proc. 555, 226 \\
2002 & 58   & S   & Sandage, A. 2002, AJ 123, 1179 \\
2002 & 59.2 & T   & Tammann, G.\,A., et~al. 2002, in A New Era in Cosmology, ASP Conf. Proc. 283, 258 \\
2002 & 56.9 & T   & Tammann, G.\,A., \& Reindl, B. 2002, in The Cosmological Model, XXXVIIth Moriond Ap. Meeting, 13 \\
\hline
\noalign{\smallskip}
   \multicolumn{4}{c}{(b) various methods, {\em not\/} corr. for bias [$\circ$]} \\
\noalign{\smallskip}
\hline
1972 & 100  & deV & de Vaucouleurs, G. 1972, in External Galaxies and Quasi-Stellar Objects, IAU Symp. 44, 353 \\
1976 & 75   & deV & de Vaucouleurs, G. 1976, ApJ 205, 13 \\
1977 & 85   & deV & de Vaucouleurs, G. 1977, in Redshifts and the Expansion of the Universe, 301 \\
1978 & 95   & deV & de Vaucouleurs, G. 1978, in The Large Scale Structure of the Universe, IAU Symp. 79, 205 \\
1981 & 96   & deV & de Vaucouleurs, G., \& Peters, W.\,L. 1981, ApJ 248, 395 \\
1986 & 109  & deV & de Vaucouleurs, G., \& Peters, W.\,L. 1986, ApJ 303, 19 \\
1986 & 99   & deV & de Vaucouleurs, G., \& Corwin, H.\,G. 1986, ApJ 308, 487 \\
1986 & 95   & deV & de Vaucouleurs, G. 1986, in Galaxy distances and deviations from universal expansion, eds. B.\,F. Madore \& R.\,B. Tully, (Dordrecht: Reidel), 1 \\
1993 & 85   & deV & de Vaucouleurs, G. 1993, ApJ 415, 10 \\
1993 & 90   & Tu  & Tully, R.\,B. 1993, in Proc. Nat. Acad. Sci. 90, 4806 \\
1997 & 81   & Gz  & Gonzales, A.\,H., \& Faber, S.M. 1997, ApJ 485, 80 \\
1997 & 73   & M   & Mould, J.\,R., et~al. 1997, in The Extragalactic Distance Scale, ed. M. Livio (Cambridge Univ. Press), 158 \\
2001 & 72   & Fr  & Freedman, W.\,L., et~al. 2001, ApJ 553, 47 \\
\hline
\noalign{\smallskip}
   \multicolumn{4}{c}{(c) SNe\,Ia [$\diamond$]} \\
\noalign{\smallskip}
\hline
1982 & 50   & ST  & Sandage, A., \& Tammann, G.\,A. 1982, ApJ 256, 339 \\
1988 & 59   & Br  & Branch, D. 1988, in The Extragalactic Distance Scale, ASP Conf. Ser. 4, 146 \\
1990 & 46.5 & TL  & Tammann, G.\,A., \& Leibundgut, M. 1990, A\&A 236, 9 \\
1994 & 52   & SN  & Saha, A., et~al. 1994, ApJ 425, 14 \\
1995 & 52   & SN  & Saha, A., et~al. 1995, ApJ 438, 8 \\
1995 & 56.5 & SN  & Tammann, G.\,A., \& Sandage, A. 1995, ApJ 452, 16 \\
1995 & 71   & Pi  & Pierce, M.\,J., \& Jacoby, G.H. 1995, AJ 110, 2885 \\
1996 & 56.5 & SN  & Saha, A., et~al. 1996, ApJ 466, 55 \\
1996 & 63.1 & H   & Hamuy, M., et~al. 1996, AJ 112, 2398 \\
1997 & 56   & SN  & Saha, A., et~al. 1997, ApJ 486, 1 \\
1999 & 60   & SN  & Saha, A., et~al. 1997, ApJ 522, 802 \\
1999 & 62.9 & TB  & Tripp, R., \& Branch, D. 1999, ApJ 525, 209 \\
1999 & 63.9 & Su  & Suntzeff, N.\,B., et~al. 1999, ApJ 500, 525 \\
1999 & 63.3 & Ph  & Phillips, M.\,M. 1999, AJ 118, 1766 \\
1999 & 64.4 & J   & Jha, S., et~al. 1999, ApJS 125, 73 \\
2000 & 68   & G   & Gibson, B.\,K., et~al. 2000, ApJ 529, 723 \\
2000 & 58.5 & SN  & Parodi, B.\,R., et~al. 2000, ApJ 540, 634 \\
2001 & 71   & Fr  & Freedman, W.\,L., et~al. 2001, ApJ 553, 47 \\
2001 & 58.7 & SN  & Saha, A., et~al. 2001, ApJ 562, 314 \\
2004 & 71   & A   & Altavilla, G., et~al. 2004, MNRAS 349, 1344 \\
2005 & 73   & R   & Riess, A.\,G., et~al. 2005, ApJ 627, 579 \\
2006 & 62.3 & SN  & Tammann, G.\,A., et~al., ApJ, to be published \\
\hline
\noalign{\smallskip}
   \multicolumn{4}{c}{(d) Tully-Fisher, corr. for bias [$\blacktriangle$]} \\
\noalign{\smallskip}
\hline
1976 & 50   & ST  & Sandage, A., \& Tammann, G.\,A. 1976, ApJ 210, 7 \\
1997 & 55   & Th  & Theureau, G., et~al. 1997, A\&A 322, 730 \\
1999 & 58   & F   & Federspiel, M. 1999, Ph.D. Thesis, Univ. of Basel \\
1999 & 53   & E   & Ekholm, T., et~al. 1999, A\&A 347, 99 \\
2000 & 55   & Th  & Theureau, G. 2000, in XIX Texas Symposium, eds. E. Aubourg et al. Mini-Symp. 13/12 \\
2002 & 65   & He  & Hendry, M.\,A. 2002, in New Era in Cosmology, eds. T. Shanks, \& N. Metcalfe, ASP Conf. Ser. 283, 258 \\
\hline
\noalign{\smallskip}
   \multicolumn{4}{c}{(e) Tully-Fisher, {\em not\/} corr. for bias [$\vartriangle$]} \\
\noalign{\smallskip}
\hline
1977 & 82   & Tu  & Tully, R.\,B., \& Fisher, J.\,R. 1977, in Redshifts and the Expansion of the Universe, 95 \\
1980 & 95   & Aa  & Aaronson, M., et al. 1980, ApJ 239, 12 \\
1984 & 91   & Bo  & Bothun, G.\,D., et al. 1984, ApJ 278, 475 \\
1986 & 90   & Aa  & Aaronson, M., et al. 1986, ApJ 302, 536 \\
1988 & 85   & Pi  & Pierce. M.\,J., \& Tully, R.\,B. 1988, ApJ 330, 579 \\
1988 & 85   & H   & Huchra, J.\,P. 1988, in The Extragalactic Distance Scale, ASP Conf. Ser. 4, 257 \\
1994 & 86   & Pi  & Pierce, M.\,J. 1994, ApJ 430, 53 \\
1997 & 70   & Gi  & Giovanelli, R. 1997, in The Extragalactic Distance Scale, ed. M. Livio (Cambridge Univ. Press), 113 \\
2000 & 77   & Tu  & Tully, R.\,B., \& Pierce, M.\,J. 2000, ApJ 533, 744 \\
2000 & 81   & R   & Rothberg, B., et~al. 2000, ApJ 533, 781 \\
2000 & 71   & Sk  & Sakai, S., et~al. 2000, ApJ 529, 698 \\
\hline
\noalign{\smallskip}
   \multicolumn{4}{c}{(f) D$_{n}$-$\sigma$, fundamental plane  [$+$]} \\
\noalign{\smallskip}
\hline
1987 & 67   & D   & Dressler, A. 1987, ApJ 317, 1 \\
1999 & 52   & F   & Federspiel, M. 1999, Ph.D. Thesis, Univ. of Basel \\
2000 & 78   & K   & Kelson, D.\,D., et~al. 2000, ApJ 529, 768 \\
\hline
\noalign{\smallskip}
   \multicolumn{4}{c}{(g) globular clusters[$\times$]} \\
\noalign{\smallskip}
\hline
1979 & 80   & Hn  & Hanes, D.\,A. 1979, MNRAS 188, 901 \\
1988 & 61   & Hs  & Harris, W.\,E. 1988, in The Extragalactic Distance Scale, ASP Conf. Ser. 4, 231 \\
1993 & 85   & deV & de Vaucouleurs, G. 1993, ApJ 415, 33 \\
1995 & 78   & W   & Whitmore, B.\,C., et~al. 1995, ApJ 454, 73 \\
1996 & 68   & B   & Baum, W.\,A., et~al. 1996, A\&AS 189, 1204 \\
1997 & 82   & W   & Whitmore, B.\,C. 1997, in The Extragalactic Distance Scale, ed. M. Livio (Cambridge Univ. Press), 254 \\
2000 & 69   & K   & Kavelaars, J.\,J., et~al. 2000, ApJ 533, 125 \\
\hline
\noalign{\smallskip}
   \multicolumn{4}{c}{(h) planetary nebulae [{\footnotesize $\circ$}]} \\
\noalign{\smallskip}
\hline
1990 & 87   & Ja  & Jacoby, G.\,H., et~al. 1990, ApJ 356, 332 \\
1991 & 77   & Bo  & Bottinelli, L., et~al. 1991, ApJ 252, 550 \\
1993 & 75   & Mc  & McMillan, R., et~al. 1993, ApJ 416, 62 \\
2002 & 78   & Ci  & Ciardello, R., et~al. 2002, ApJ 577, 31 \\
\hline
\noalign{\smallskip}
   \multicolumn{4}{c}{(i) surface brightness fluctuations [{\scriptsize $\circ$}]} \\
\noalign{\smallskip}
\hline
1989 & 88   & To  & Tonry, J.\,L., et~al., 1989, ApJ 346, 57 \\
1997 & 81   & To  & Tonry, J.\,L. 1997, in The Extragalactic Distance Scale, ed. M. Livio (Cambridge Univ. Press), 297 \\
1998 & 82   & La  & Lauer, T.\,R., et~al. 1998, ApJ 499, 577 \\
1999 & 74   & Bl  & Blakeslee, J.\,P., et~al. 1999, ApJ 527, 73 \\
1999 & 87   & Je  & Jensen, J.\,B., et~al. 1999, ApJ 510, 71 \\
2000 & 77   & To  & Tonry, J.\,L., et~al. 2000, ApJ 530, 625 \\
2001 & 73   & Aj  & Ajhar, E.\,A., et~al. 2001, ApJ 559, 584 \\
\end{longtable}
\normalsize

\section{HST and \boldmath{$H_{0}$}} 
\label{sec:05}
With the advent of HST two major campaigns were started for the
determination of $H_{0}$.

\subsection{The HST Key Project on the Extragalactic Distance Scale} 
\label{sec:05:1}
The original program was to observe Cepheids in many inclined spirals
in order to provide a calibration for the $I$-band TF relation
\citep{Aaronson:Mould:86}; at the time the authors still favored a
value of $H_{0}=90$. Later the Cepheid distances were planned 
\citep{Kennicutt:etal:95} to also calibrate the LF of PNe and
the expanding-atmosphere parallaxes of SNe\,II, of novae, and of the
peak of the LF of GCs. Surprisingly the authors made only cursory
reference to the problem of Malmquist bias. The program team,
consisting of 26 collaborators provided 19, i.e.\ almost half of all
published Cepheid distances. The distances were based on the P-L
relation of 22 LMC Cepheids and a zero-point set at $(m-M)^{0}_{\rm
  LMC}=18.50$ \citep{Madore:Freedman:91}. In a first summary paper
\citet{Mould:etal:00} concluded from the TF and SBF methods, from
SNe\,Ia and, now also from the FP method that $H_{0}=68\pm6$, if they
made allowance for high-metallicity, (long-period) Cepheids being
somewhat brighter than their LMC counterparts. Unfortunately
\citet{Freedman:etal:01} raised the result to $H_{0}=72\pm8$ on the
basis of an interim P-L relation \citep{Udalski:etal:99} which is now
untenable. 

\subsection{The HST Project for the Luminosity-Calibration of SNe\,Ia} 
\label{sec:05:2}
A small group of astronomers (A.~Saha, F.\,D.~Macchetto, N.~Panagia,
I, and A.~Sandage as PI) proposed to observe Cepheids with HST in
galaxies which had produced a well observed SN\,Ia. The results for 8
galaxies were published; 4 additional ones came from external sources
\citep{Turner:etal:98,Tanvir:etal:99,Macri:etal:01,Riess:etal:05}.
Two out of the 12 SNe\,Ia are spectroscopically peculiar and were
excluded, leaving 10 Cepheid distances for the calibration of normal
SNe\,Ia. The program has only recently been completed because (1) the
WFPC2 on HST was to be recalibrated \citep{Saha:etal:05}, and (2)
unexpected complications were found with the P-L relation of Cepheids
(see below Sec.~\ref{sec:05:2:2}). The route to $H_{0}$ was described
in five papers 
\citep{Tammann:etal:03,Sandage:etal:04,Sandage:etal:06,Reindl:etal:05,Saha:etal:05},
of which only a summary is given here.

\subsubsection{The Hubble diagram of SNe\,Ia} 
\label{sec:05:2:1}
The first Hubble diagram of SNe\,Ia was shown by \citet{Kowal:68}. Its
large dispersion was steadily decreased by subsequent authors. By 1979
SNe\,Ia had emerged as so reliable standard candles that it could be
proposed to observe them at large redshifts ($z\ga0.5$) for a
determination of $\Lambda$ \citep{Tammann:79}. It is well known that
this has become possible since; how much easier must it be to use
SNe\,Ia at small redshifts for a determination of $H_{0}$! -- if only
their luminosity calibration is realized.

\begin{figure}[t]
\centering
\resizebox{0.55\textwidth}{!}{\includegraphics{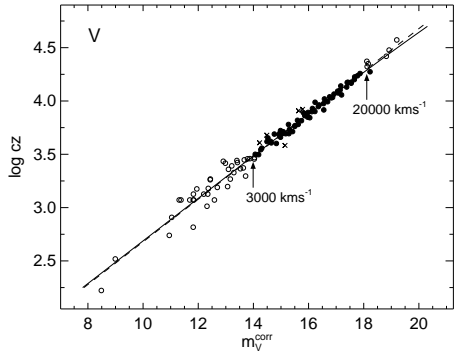}}
\caption{The Hubble diagram in $V$ of 111 normal SNe\,Ia. The objects
  outside the indicated velocity range are shown as open symbols; at
  low velocities the scatter increases because of the influence of
  peculiar velocities. The slightly curved Hubble line for
  $\Omega_{\rm M}=0.3$, $\Omega_{\Lambda}=0.7$ is a fit to only the
  black symbols; the crosses are not considered for the fit. The
  dashed line holds for an $\Omega_{\rm T}=0$ Universe.}
\label{fig:10}
\end{figure}

     There are now 124 SNe\,Ia nearer than $30\,000\kms$ with known 
$B$, $V$ and in most cases $I$ magnitudes at maximum as well as
decline rates $\Delta m_{15}$ (the decline in mag over the first 15
days past $B_{\max}$). Excluding 13 spectroscopically peculiar objects
leaves 111 normal SNe\,Ia. Their magnitudes are corrected for Galactic
and internal absorption \citep{Reindl:etal:05}. The internal reddening
is determined by adopting the intrinsic colors $(B\!-\!V)^{0}$ and
$V\!-\!I)^{0}$  -- and their non-negligible dependence on $\Delta
m_{15}$ -- from 21 SNe\,Ia in (almost) dust-free E/S0 galaxies. The
absorption-corrected absolute magnitudes $M^{0}_{BVI}$, calculated
from velocity distances, correlate with the Hubble type of the parent
galaxy, SNe\,Ia in early-type galaxies being fainter. This dependence
on Hubble type can empirically be removed by normalizing the
magnitudes to a standard value of the decline rate, say $\Delta
m_{15}=1.10$. Also the slight dependence of the luminosity on
$(B\!-\!V)^{0}$ is removed by normalizing to the color at $\Delta
m_{15}=1.1$ [$(B\!-\!V)^{0}_{1.1}=-0.024$]. The resulting magnitudes
$m^{\rm corr}_{BVI}$ can be plotted in a Hubble diagram; as an example
$m^{\rm corr}_{V}$ is shown in Fig.~\ref{fig:10}. A fiducial sample
of 62 normal SNe\,Ia with $3000<v_{\rm CMB}<20\,000\kms$, i.e.\ in the
ideal range to calibrate the large-scale value of $H_{0}$, define a
Hubble line of 
\begin{equation}
     \log v= 0.2m^{\rm corr}_{\lambda} + C_{\lambda},
\label{eq:06}
\end{equation}
with $C_{B}=0.693\pm0.004$, $C_{V}=0.688\pm0.004$,
$C_{I}=0.637\pm0.004$. The solution for the intercept $C_{\lambda}$ is
very robust against chosing different SN subsets
\citep[see][Table~9]{Reindl:etal:05}. The small scatter of
$\sigma_{\rm m}=0\fm15$ -- smaller than for any other known individual
objects -- makes SNe\,Ia ideal standard candles. In fact much of the
scatter is driven by errors of the internal absorption correction,
because the 21 SNe\,Ia in E/S0's have a scatter in $I$ of only
$0\fm10$! Transforming eq.~(\ref{eq:06}) yields
\begin{equation}
     \log H_{0}= 0.2M^{\rm corr}_{\lambda} + C_{\lambda}.
\label{eq:07}
\end{equation}
In order to obtain $H_{0}$ it remains ``only'' to calibrate 
$M^{\rm corr}_{\lambda}$ for some nearby SNe\,Ia with known Cepheid
distances. It may be noted that the error of $C_{\lambda}$ is so
small, that the statistical error of $H_{0}$ will essentially depend
on only the error of $M^{\rm corr}_{\lambda}$.

\subsubsection{Cepheid distances of galaxies with SNe\,Ia} 
\label{sec:05:2:2}
The determination of Cepheid distances has become much more
complicated since it has been realized that the P-L relation is {\em
  not\/} universal. In particular the relations in the Galaxy and in
LMC are significantly different \citep{Tammann:etal:02}. The Galactic
P-L relation in $BVI$ is quite well defined by 33 Cepheids in open
clusters (with a zero-point at $(m-M)^{0}_{\rm Pleiades}=5\fm61$)
{\em and\/} 36 Cepheids with moving-atmosphere (BBW) parallaxes by
\citet{Fouque:etal:03} and a few others 
\citep[see also][]{Ngeow:Kanbur:04}. The P-L relation in $BVI$ 
of LMC rests on 593 very well observed Cepheids from the OGLE program
\citep{Udalski:etal:99} and 97 bright Cepheids from various sources as
well as an adopted zero-point of $(m-M)^{0}_{\rm
  LMC}=18.54$. Long-period Galactic Cepheids with a mean metallicity
of [O/H]$\;=8.60$ are {\em brighter\/} than their LMC counterparts with
[O/H]$\;=8.36$. The details of the two different P-L relations are laid
out in \citet{Tammann:etal:03} and \citet{Sandage:etal:04}. An
important feature of the LMC P-L relation is that its slope breaks at
$P=10^{\rm d}$ \citep[see also][]{Ngeow:etal:05}, which is not seen in
the Galaxy (and SMC). 

\begin{figure}[t]
\centering
\resizebox{0.5\textwidth}{!}{\includegraphics{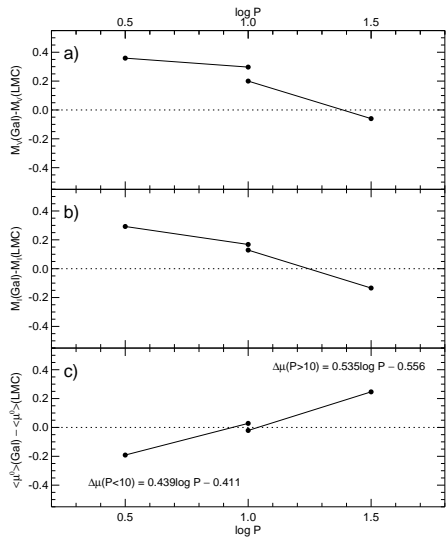}}
\caption{a) The difference between the absolute $V$ magnitude of a
  Cepheid determined once from the Galactic and once from the LMC P-L
  relation. b) Same for $I$ magnitudes. c) The distance difference
  from eq~(\ref{eq:08}) if once $M_{V}$ and $M_{I}$ are taken from the
  Galactic and once from the LMC P-L relation. All differences are
  plotted in function of $\log P$.}
\label{fig:11}
\end{figure}

     The crux is that the two different P-L relations yield two
different distances for every galaxy. The problem is aggrevated when
only $V$ and $I$ magnitudes are used, as in the case of the HST
observations, to determine the true distance modulus {\em and\/} the
mean internal absorption of the Cepheids. In that case the true and
apparent moduli are connected by
\begin{equation}
     (m-M)^{0}= 2.52(m-M)_{I} - 1.52(m-M)_{V}.
\label{eq:08}
\end{equation}
(The coefficients depend on the adopted absorption-to-reddening ratio
${\cal R}$). Fig.~\ref{fig:11} shows how the magnitudes $M_{V}$
and $M_{I}$ as well as the true moduli differ in function of period
when once the Galactic P-L relation is used and once the one from
LMC. The moduli of the former are larger by up to $0\fm25$ at $\log
P=1.5$. 

     Only a small part of the P-L relation differences can be
explained as the line blanketing effect of the metals, but the main
effect is that LMC Cepheids are hotter than Galactic Cepheids at given
period or given luminosity. The reason is unknown at present. But
\citet{Saha:etal:05} have made the {\em assumption\/} that the whole
difference is a metallicity effect. Consequently they have derived
Cepheid distances of 37 galaxies by interpolating (and slightly
extrapolating) their distances from Galactic and LMC P-L relations
according to their metallicity as measured by [O/H]. The ensuing
metallicity corrections are somewhat larger than proposed by
\citet{Kennicutt:etal:98} and \citet{Sakai:etal:04}, but they are
justified by several comparisons with external data; for instance the
adopted Cepheid distances of nine galaxies for which also independent,
metal-insensitive TRGB distances are available \citep{Sakai:etal:04}
show no systematic trend with [O/H]. Also the resulting SN\,Ia
luminosities do not show a significant correlation with the
metallicity of their parent galaxies. Finally it may be noted thar the
metal-rich (inner) and metal-poor (outer) Cepheids in M\,101 give the
same distance to within $0\fm01$.

\subsubsection{The value of \boldmath{$H_{0}$}} 
\label{sec:05:2:3}
As mentioned before there are 10 normal SNe\,Ia in galaxies with
Cepheid distances. The absorption-corrected, normalized magnitudes
$m^{\rm corr}_{BVI}$ of these SNe\,Ia are derived in exactly the same
way -- and this is an important point -- as for the distant SNe\,Ia
which define the Hubble diagram in Fig.~\ref{fig:10}
\citep{Reindl:etal:05}. The metallicity-corrected distances of the 10
SN\,Ia parent galaxies are derived by \citet{Saha:etal:05}. Combining
the magnitudes $m^{\rm corr}_{BVI}$ with the corresponding distances
yields immediately the absolute magnitudes; the weighted means become
$M^{\rm corr}_{B}=-19.49\pm0.04$, $M^{\rm corr}_{V}=-19.46\pm0.04$,
and $M^{\rm corr}_{I}=-19.22\pm0.05$. By inserting the absolute
magnitudes with their appropriate intercepts $C_{\lambda}$
(eq.~\ref{eq:06}) into eq.~(\ref{eq:07}) one finds
$H_{0}(B)=62.4\pm1.2$, $H_{0}(V)=62.4\pm1.5$, and
$H_{0}(I)=62.1\pm1.4$, or the mean for scales of the order of
$20\,000\kms$
\begin{equation}
     H_{0}=62.3\pm1.3\quad (\mbox{random error}).
\label{eq:09}
\end{equation}
The systematic error of this result has been discussed in some detail
by \citet{Sandage:etal:06} and estimated to be $\pm5.3$, most of which
is caused by the non-uniqueness of the P-L relation of Cepheids and
the closely related question of the metallicity correction of Cepheid
distances. 

\section{The Local Value of \boldmath{$H_{0}$} and the Random Motions
  of Field Galaxies}
\label{sec:06}
\begin{figure}[t]
\centering
\resizebox{1.0\textwidth}{!}{\includegraphics{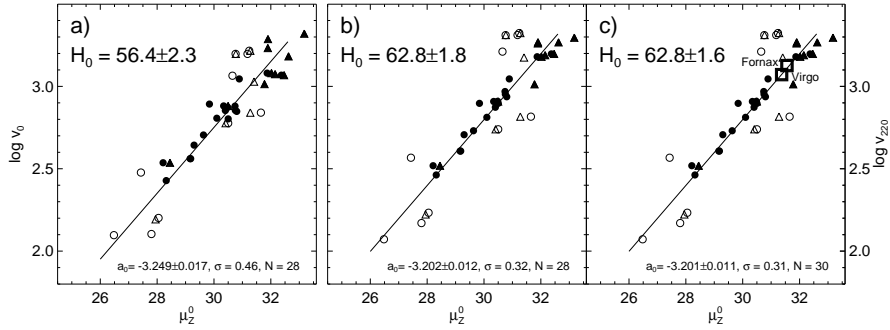}}
\caption{The local distance-calibrated Hubble diagram of 34 galaxies
  with $v_{220}<2000\kms$ for which 27 Cepheid distances (dots)
  and 16 SN\,Ia distances (triangles) are available. Galaxies
  within $25^{\circ}$ from the Virgo cluster or with $(m-M)^{0}<28.2$
  are shown as open symbols. The Hubble line is a fit to only the
  closed symbols. a) using velocities $v_{0}$ corrected to the
  barycenter of the Local Group. b) using velocities $v_{220}$
  corrected for Virgocentric infall. c) same as b) but with the Virgo
  and Fornax clusters added.}
\label{fig:12}
\end{figure}

For a number of ``local'' galaxies with $v_{220}<2000\kms$ Cepheid
and/or SN\,Ia distances are available. Excluding members of the Virgo
and Fornax clusters and four nearby galaxies with $(m-M)^{0}<28.2$
leaves 34 galaxies with at least one distance determination. Their
distance-calibrated Hubble-diagram shows a very large scatter of
$\sigma_{\rm m}=1\fm0$, which can only be due to peculiar
velocities. It is reduced to $\sigma_{\rm m}=0\fm46$ if the 12
galaxies are omitted whose distance from the Virgo cluster (M\,87) is
$<\!25^{\circ}$ (Fig.~\ref{fig:12}a). Clearly a region of 
$25^{\circ}$ ($\sim\!8\;$Mpc) radius about the Virgo cluster is
characterized by much larger turbulent motions than the ``normal''
field. The scatter in the field is further reduced to  $\sigma_{\rm
  m}=0\fm32$ if the velocities $v_{0}$ (corrected to the centroid of the
Local Group; \citealt{Yahil:etal:77}) are replaced by $v_{220}$. The
$v_{220}$ velocities are corrected for a selfconsistent Virgocentric
infall model with a local Virgocentric velocity vector of $220\kms$
\citep{Yahil:etal:80,Tammann:Sandage:85,Kraan-Korteweg:86}. This model
finds here support from the unexpectedly small scatter in 
Fig.~\ref{fig:12}b, where the $v_{220}$ velocities are used.

     In Fig.~\ref{fig:12}c also the Virgo and Fornax clusters are
plotted with their mean Cepheid and SN\,Ia distances ($\langle
m-M\rangle^{0}_{\rm Virgo}=31.47\pm0.16$, $\langle v_{220}\rangle
=1179\kms$ and $\langle m-M\rangle^{0}_{\rm Fornax}=31.56\pm0.13$,
$\langle v_{220}\rangle =1338\kms$; see
\citealt{Sandage:etal:06}). They fit on the Hubble line well within
the errors.

     The resulting value of $H_{0}(\mbox{local})=62.8\pm1.6$ is
undistinguishable from the large-scale value. This does not mean that
the expansion is blind toward density fluctuations, because the
gravitational effect of the Virgo cluster complex has been eliminated
by subtracting the Virgocentric flow model.

     The small scatter of $0\fm32$ in Fig.~\ref{fig:12}b of the
field galaxies outside the $25^{\circ}$ circle puts strong upper
limits on the size of the peculiar motions, i.e.\ $\partial v/v=0.16$
even without allowing for distance errors. The typical peculiar
velocity of a galaxy at say $1000\kms$ is therefore $<160\kms$. --
Also the peculiar motions of more distant field galaxies are
restricted by SNe\,Ia. The 20 SNe\,Ia in E/S0 galaxies (and hence
little internal absorption) with $5000<v_{\rm CMB}<20\,000\kms$
scatter about the Hubble line, as stressed before, by only $0\fm10$
(in $I$ magnitudes, \citealt{Reindl:etal:05}). Some of this scatter
must be due to photometric errors and to the intrinsic dispersion of
the normalized SN\,Ia magnitudes; $\partial v/v=0.05$ or $v_{\rm
  pec}=300\kms$ at a distance of $6000\kms$ are therefore generous
upper limits.

\section{Concluding Remarks}
\label{sec:07}
In general astronomical distances depend on objects whose distances
are already known and ultimately, with a few exceptions, on
trigonometric parallaxes and hence on the AU. But methods of
determining distances from the physics or geometry of some objects,
without recourse to any other astronomical distance, are gaining
increasing weight. Already the moving-atmosphere (BBW) method
contributes to the calibration of the Galactic P-L relation of
Cepheids. The single, intrinsicly accurate water maser distance of
NGC\,4258 \citep{Herrnstein:etal:99} does not yet suffice for an
independent calibration of the P-L relation
\citep[see][]{Saha:etal:05}. The recently improved
expanding-atmosphere distance of SN\,II 1999em \citep{Baron:etal:04}
agrees well with the Cepheid distance of its parent galaxy NGC\,1637. 
\citet{Nadyozhin:03} plateau-tail method for SNe\,IIP yields
$H_{0}=55\pm5$ on the assumption that the $^{56}$Ni mass equals the
explosion energy. Models of SNe\,Ia yield $M_{\rm bol}\approx
M_{V}=-19.5$ \citep[][for a review]{Branch:98} in fortuitous agreement
with the empirical value of $M_{V}=-19.46$. 

     Much promise to determine $H_{0}$ accurately lies in the
Sunyaev-Zeldovich (SZ) effect and in gravitationally lensed quasars;
extensive work has gone into both methods. The SZ effect yields
typical values of $H_{0}=60\pm3$, yet the systematic error is still
$\sim\!\pm18$ (\citealt{Carlstrom:etal:02} for a review; see also
e.g.\ \citealt{Udomprasert:etal:04}; \citealt{Jones:etal:05}). Results
from lensed quasars lie still in a wide range of $48<H_{0}<75$
\citep[e.g.][]{Saha:03,Koopmans:etal:03,Kochanek:Schechter:04,York:etal:05}.

     A strong driver to determine $H_{0}$ as accurately as possible
comes from the CMB. The interpretation of its Fourier spectrum
depends on at least twelve free parameters, several of which cannot be
determined elsewhere. A simultaneous solution for all twelve
parameters yields $H_{0}=66\pm7$ \citep{Rebolo:etal:04}. It would
bring important progress for the understanding of the CMB spectrum if
an independently determined value of $H_{0}$ could be used as a
reliable prior.

     The value of $H_{0}=62.3$ corresponds to an expansion age of
$15.1\;$Gyr in a flat $\Lambda$CDM model with
$\Omega_{\Lambda}=0.7$. This gives a sufficient time 
frame for the oldest ages in the Galaxy. Stellar-evolution models give
for M\,107 $14.0\pm2.8$ \citep{Chaboyer:etal:00} and for
M\,92 $13.5\;$Gyr \citep{VandenBerg:etal:02}. Models of the
chemical evolution of the Galaxy yield ages of the actinides of
$12.4-14.7\;$Gyr \citep{Thielemann:etal:87}
or a U/Th age of $14.5\pm2.5$ \citep{Dauphas:05}.
The emphasis has shifted over the last years to the Th/Eu dating of
ultra-metal-poor giants.
Typical results lie between $14.2\pm3.0$ and $15.6\pm4.0\;$Gyr 
\citep{Cowan:etal:99,Westin:etal:00,Truran:etal:01,Sneden:etal:03}.
The ages are to be increased by the gestation time of the relevant
objects; some galaxies may also have started their star formation
before the Galaxy did. The present age determinations are not yet
sufficiently accurate to set stringent limits on $H_{0}$, but the
essential point is that none of these ages are significantly larger
than allowed for by the expansion age.

\newpage

\noindent
{\em Acknowledgement.} 
My sincere thanks go to the Astronomische Gesellschaft which gave me
the honorable and joyful opportunity for this report. I want to thank
also my two honored teachers, Wilhelm Becker, himself a
Karl-Schwarzschild Lecturer, and Allan Sandage, with whom I had the
chance to collaborate for 43 years. Many other collaborators, too
numerous to be named here, have helped me with their ideas and
knowledge and friendship. Without the unbureaucratic support of the
Swiss National Science Foundation during more than three decennia my
scientific life would have been poorer.



\end{document}